\documentstyle[prd,aps,floats]{revtex}

\begin{document}
\preprint{archive/ddmmyy}
\draft

\input epsf
\renewcommand{\topfraction}{0.8}
\renewcommand{\bottomfraction}{0.8}
\twocolumn[\hsize\textwidth\columnwidth\hsize\csname 
@twocolumnfalse\endcsname

\title{Generalized Compactification and 
Assisted Dynamics of Multi--Scalar Field Cosmologies}
\author{Anne M. Green and James E. Lidsey} 
\address{ Astronomy Unit, School of Mathematical Sciences,
Queen Mary and Westfield College,\\ Mile End Road, London, E1
4NS,~~U.~~K.  } 
\date{\today} 
\maketitle
\begin{abstract}
Cosmological models arising from a generalized compactification of
Einstein gravity are derived.  It is shown that a redefinition of the
moduli fields reduces the system to a set of massless fields and a
single field with a single exponential potential, independent of the
background spacetime.  This solution is the unique late--time
attractor for an arbitrary spacetime dimensionality. We find that if
the number of dimensions is greater than or equal to seven, the scalar
fields dominate a relativistic fluid and therefore constitute a
potential `moduli' problem. 

\end{abstract}


\vskip2pc]

In standard Kaluza--Klein compactifications it is assumed that the
higher--dimensional fields are independent of the compactified
coordinates \cite{standard}.  However, when the action exhibits a
global symmetry, consistent lower--dimensional theories may be derived
from the more general ansatz of Scherk and Schwarz \cite{SchSch}.  The
simplest example is the compactification on a circle of a theory
containing a massless axion field. Since the field arises in the
action only through its derivative, the global symmetry corresponds to
a linear shift in its value.  A consistent compactification is then
possible if the axion has a {\em linear} dependence on the internal
variable \cite{massaxion,nkm}. The slope parameter manifests itself in
the compactified theory as an exponential self--interaction potential
of the modulus field that parametrises the radius of the internal
dimension.  There has recently been considerable interest in
generalised Scherk--Schwarz compactifications
\cite{massaxion,nkm,renewed,ortin} within the context of string
duality and M--theory \cite{Mtheory}.  This has led to a deeper
understanding of the relationships between the different string
theories.  In particular, it has been shown \cite{ortin} that the
Scherk--Schwarz compactified type IIB theory is T--dual to the massive
type IIA supergravity theory due to Romans \cite{romans}.

It is also important to study the cosmological implications of
compactifications of this type and this is the purpose of the present
paper. In general, the resulting four--dimensional actions consist of
a set of scalar fields interacting through a series of exponential
potentials. Until recently, it was widely thought that the potential
with the smallest slope would dominate the dynamics of such a system
at late times. Surprisingly, this is not the case 
\cite{lms}.  The authors of Ref. \cite{lms}
considered a specific model comprised of $m$ non-interacting scalar
fields, $\vec{\varphi} = (\varphi_1 , \varphi_2 , \ldots , \varphi_m
)$, each with an exponential potential, $V_i(\varphi_i) \propto \exp(
-\varphi_i/\sqrt{r}_i)$, where $r_i$ is a constant\footnote{In this
paper, units are chosen such that $16\pi G = \hbar = c =1$.}. The
late--time attractor for the spatially flat,
Friedmann--Robertson--Walker (FRW) cosmology is then of the form $a
\propto t^{\tilde{r}}$, where $\tilde{r} = \sum_{i=1}^m
r_i$~\cite{lms,mw}. Thus, inflation is possible, $\tilde{r} >1$, even
if the potentials are individually too steep to drive inflation $(r_i
< 1)$. This cooperative behaviour was termed `assisted'
inflation~\cite{lms}.

In view of the above developments, we derive a four--dimensional
cosmological model by employing a generalized Scherk--Schwarz
compactification of vacuum Einstein gravity. The dimensionality, $d$,
of the original theory is kept arbitrary in the
analysis, and the model corresponds to a consistent compactification
of the low--energy, vacuum limit of M--theory when $d=11$
\cite{Mtheory}. We then determine the nature of the unique, late--time
attractor for an arbitrary spacetime metric by redefining the moduli
fields in a suitable way\footnote{We assume implicitly that the
universe does not recollapse before the attractor solution becomes
relevant.}. 
The assisted dynamics between the scalar fields leads to
interesting behaviour and may result in a new moduli problem for the
early universe.

We begin by considering $(D+1)$--dimensional Einstein gravity minimally 
coupled to a massless scalar field, $\hat{\Phi}$: 
\begin{equation}
S=\int d^{D+1} x \sqrt{-\hat{g}_{D+1}} \left[ \hat{R} -\frac{1}{2} 
\left( \hat{\nabla} \hat{\Phi} \right)^2 \right] \,,
\end{equation} 
where $\hat{R}$ is the Ricci scalar of the spacetime with metric 
$\hat{g}_{MN}$ and $\hat{g} \equiv {\rm det} \hat{g}_{MN}$.
Compactification onto a circle may be parametrised in terms of the metric 
\cite{lp}
\begin{eqnarray}
\label{metric}
ds_{D+1}^2 =e^{2\alpha \varphi} ds_D^2 +e^{-2(D-2) \alpha \varphi} \left( 
dz +A_{\mu}dx^{\mu} \right)^2  \,, \\
\label{alpha}
\alpha \equiv - \frac{1}{\sqrt{2(D-1)(D-2)}} \,,
\end{eqnarray}
where $z$ represents the coordinate of the compactified dimension,
$A_{\mu}$ is the gauge potential and the numerical value of $\alpha$
is chosen to ensure that the scalar `dilaton' field, $\varphi$, is
minimally coupled to the Einstein--frame metric after
compactification. For the ansatz
\begin{equation}
\label{linearansatz}
\hat{\Phi}(x^{\mu},z) =\Phi (x^{\mu}) + m z \,,
\end{equation}
the reduced action is given by \cite{massaxion}
\begin{eqnarray}
\label{reducedaction}
S&=&\int d^D x \sqrt{-g_D} \left[ 
R -\frac{1}{2} \left( \nabla \varphi \right)^2 -\frac{1}{4} e^{a\varphi} 
F_{\mu\nu}F^{\mu\nu}  \nonumber \right.  \\ & & \left.
-\frac{1}{2} \left( {\cal{D}} \Phi \right)^2 -\frac{1}{2} 
m^2 e^{-a\varphi} \right] \,,
\end{eqnarray}
where ${\cal{D}}_{\mu} \Phi \equiv \partial_{\mu} \Phi -mA_{\mu}$,
$F_{\mu\nu} =2\partial_{[\mu } A_{\nu ]}$ is the field strength of
$A_{\mu}$ and $a\equiv \sqrt{2(D-1)/(D-2)}$.  Action
(\ref{reducedaction}) is invariant under the massive gauge
transformation $\delta \Phi =m \chi$ and $\delta A_{\mu}
=\partial_{\mu} \chi$ and this allows the vector field to gain a mass
by absorbing the axion \cite{massaxion}. The dilaton field has an
exponential self--interaction potential due to the non--trivial slope
parameter, $m$, of the higher--dimensional axion.

The dimensional reduction summarised in
Eqs.~(\ref{metric})--(\ref{reducedaction}) may be applied to
$D$--dimensional Einstein gravity minimally coupled to $m = D-4$
massless axion fields, $\vec{\Phi}$.  These fields may be interpreted
as moduli fields arising from the compactification of
$(D+m)$--dimensional, vacuum Einstein gravity on a rectilinear torus.
We may then consider a generalised dimensional reduction on a
$(D-4)$--dimensional torus, $T^m =S^1 \times S^1 \times \ldots \times
S^1$, in terms of a series of compactifications on successive circles,
$S^1$. For each $S^1$, we assume that one of the massless scalar
fields has a linear dependence on the coordinate parametrising the
circle and that the remaining fields are independent of this
coordinate. The massive vector field that arises in the dimensional
reduction may then be consistently set to zero and the process
repeated.

The result is that after compactification to four dimensions, 
the truncated action contains 
a set of $(D-4)$ minimally coupled, dilatonic
scalar fields, $\vec{\varphi}$, 
that parametrise the radii of the compactified coordinates. 
These fields couple exponentially to each other through  $(D-4)$ potentials
that originated from the higher--dimensional axions. The action 
can therefore be expressed in the form 
\begin{equation}
\label{4}
S=\int d^4 x \sqrt{-g} \left[ R 
-\frac{1}{2} \left( \nabla \vec{\varphi} \right)^2 -\frac{1}{2} 
\sum_{i=1}^{n} m_i^2 e^{-\vec{c}_i .\vec{\varphi} }\right] \,,
\end{equation}
where the constant vectors, $\vec{c}_i$, parametrise the couplings between the
fields and $m_i$ are arbitrary constants. 
The couplings are determined  by induction  
from Eqs.~(\ref{metric})--(\ref{reducedaction}) and are given by 
\begin{equation}
\label{cvectors}
\vec{c}_i \equiv \left( \underbrace{0,0, \ldots , 0}_{i-1}, (D-1-i)s_i , 
s_{i+1} , s_{i+2} , \ldots , s_{D-4} \right) \,,
\end{equation}
where  
\begin{equation}
\label{svectors}
s_i \equiv  \left[ \frac{2}{(D-1-i)(D-2-i)}\right]^{1/2} \,. 
\end{equation}
By performing linear translations on the values of the scalar 
fields, the constants, $m_i$, 
can be rescaled without loss of generality such that $m^2_i \equiv M^2/n$
for all $i$, where $M$ is a constant. 
The scalar field equations derived from action (\ref{4}) are then 
given by 
\begin{equation}
\label{field}
\nabla^2 \vec{\varphi} +\frac{M^2}{2n} \sum_{i=1}^n \vec{c}_i 
e^{-\vec{c}_i . \vec{\varphi}} =0 \,.
\end{equation}

The assisted dynamics between the scalar fields arising in action (\ref{4}) 
becomes apparent on performing an appropriate field redefinition; 
following the analysis of L\"u and Pope~\cite{lp,lp1} we rotate the
fields, $\vec{\varphi}$, with respect to a unit vector, $\vec{n}$:
\begin{equation}
\vec{\varphi} = \varphi \vec{n} +\vec{\varphi}_{\perp} \,,
\end{equation}
where $\varphi$ is a scalar function and $\vec{\varphi}_{\perp}$ is
perpendicular to $\vec{n}$: $\vec{n} . \vec{\varphi}_{\perp} =0$.  The
unit vector is chosen such that the set of vectors, $\vec{c}_i$, each
have the same projection onto it, i.e., such that
\begin{equation}
\label{definec}
\vec{c}_i.\vec{n} \equiv c \qquad \forall \qquad i \,,
\end{equation}
for some constant, $c$. 

Taking the dot product of Eq.~(\ref{field}) with respect to $\vec{n}$
then implies that
\begin{equation}
\label{boxvarphi}
\nabla^2 \varphi +\frac{cM^2}{2n}e^{-c\varphi} \sum_{i=1}^n 
e^{-\vec{c}_i. \vec{\varphi}_{\perp}} =0 \,,
\end{equation}
and substituting Eq.~(\ref{boxvarphi}) into Eq.~(\ref{field}) 
implies that 
\begin{equation}
\label{perpequation}
\nabla^2 \vec{\varphi}_{\perp} +\frac{M^2}{2n}e^{-c\varphi} \left[ 
\sum_{i=1}^n \vec{c}_i e^{-\vec{c}_i. \vec{\varphi}_{\perp}} -
c \vec{n} \sum_{i=1}^n e^{-\vec{c}_i.\vec{\varphi}_{\perp}} \right] =0 \,.
\end{equation}
It follows that if the unit vector satisfies the constraint \cite{lp}
\begin{equation}
\label{definen}
\vec{n} = \frac{1}{cn} \sum_{i=1}^n
\vec{c}_i \,,
\end{equation}
a consistent solution to Eqs.~(\ref{boxvarphi}) and (\ref{perpequation}) 
is given by 
\begin{equation}
\label{atcond}
 \vec{c}_i . \vec{\varphi}_{\perp} =0  \qquad \forall \qquad i  \,,
\end{equation}
so that 
\begin{eqnarray}
\label{fieldvarphi}
\nabla^2 \varphi +\frac{1}{2} cM^2 e^{-c \varphi} = 0 \,,  \\
\label{fieldvec}
\nabla^2 \vec{\varphi}_{\perp} =0 .
\end{eqnarray}
Eqs.~(\ref{fieldvarphi}) and (\ref{fieldvec}) 
may be derived from the action
\begin{eqnarray}
\label{singleaction}
S=\int d^4 x \sqrt{-g} \left[ R -\frac{1}{2} \left( \nabla 
\vec{\varphi}_{\perp} \right)^2 \right. \nonumber \\
\left. 
-\frac{1}{2} \left( \nabla \varphi \right)^2 
-\frac{1}{2} M^2 e^{-c \varphi} \right] 
\end{eqnarray}
with the fields perpendicular to $\vec{n}$ behaving as 
massless scalar fields.  

The numerical value of the coupling, $c$, is evaluated by taking the
dot product of Eq.~(\ref{definen}) with respect to the vector
$\vec{c}_j$.  We find that
\begin{equation}
\label{cM}
c^2 =\frac{1}{n} \sum_{i=1}^n M_{ij} \,,
\end{equation}
where the elements of the symmetric $n \times n$ matrix,  
$M_{ij}$, are determined by the dot products of the 
couplings between the fields:
\begin{equation}
\label{defineM}
M_{ij} \equiv \vec{c}_i . \vec{c}_j \,.
\end{equation}
Since Eq.~(\ref{cM}) is {\em independent} of $j$, the 
couplings must be related in a certain way. 
For the dimensionally reduced model we have derived, 
Eqs.~(\ref{cvectors}) and 
(\ref{svectors}) imply that 
\begin{eqnarray}
\label{csum}
\vec{c}_i . \vec{c}_j  = \left\{ \begin{array}{ll}
\frac{2}{D -i-2} + \sum_{k=i+1}^{D-4} \frac{2}{( D-2-k)
                       (D-1-k)},  \\
\frac{2(D-s-1)}{(D-s-2)} + \sum_{k=s+1}^{D-4} \frac{2}{( D-2-k)
                       (D-1-k)}, 
                             \end{array}
\right. 
\end{eqnarray}
where $s=\max \{i , j \}$ and the top (bottom) 
line corresponds to $i=j$ $(i\ne j)$. Employing the relationship 
\begin{equation}
\sum_{k=1}^{n} \frac{1}{[kq+(r-q)][kq+r]} \equiv \frac{n}{r(r+nq)} \,,
\end{equation}
and substituting Eq.~(\ref{csum}) into Eq.~(\ref{defineM}) then yields 
the remarkably simple result:
\begin{equation}
\label{Mcomponent}
M_{ij} =2 \delta_{ij} +1 \,.
\end{equation}
Hence, Eq.~(\ref{cM}) is satisfied and implies that $c^2 = (D-2)/(D-4)$. 

Thus far, we have established that Eq.~(\ref{singleaction}) is a
consistent truncation of action (\ref{4}), subject to the constraints
(\ref{definec}) and (\ref{definen}). The question that now arises is
whether it is also the late--time attractor for the general system.
This can be determined by taking the dot product of
Eq.~(\ref{perpequation}) with $\vec{c}_j$ and deriving an effective
equation of motion for the fields $\vec{c}_j .
\vec{\varphi}_{\perp}$:
\begin{eqnarray}
\label{cvarphieqn}
\nabla^2 \left( \vec{c}_j . \vec{\varphi}_{\perp} \right)
- \frac{M^2}{2n} e^{-c \varphi} \left[ 
\left( c^2 -\vec{c}_j .\vec{c}_j \right) e^{-\vec{c}_j. 
\vec{\varphi}_{\perp}}  \right.  \nonumber \\ \left.
 -\sum_{i\ne j}^n 
\left( c^2 -\vec{c}_i . \vec{c}_j \right)
e^{-\vec{c}_i . \vec{\varphi}_{\perp}} \right] =0 .
\end{eqnarray}
We now define the scalar field $y_{jk} \equiv \vec{c}_j.
\vec{\varphi}_{\perp} - \vec{c}_k . \vec{\varphi}_{\perp}$. 
Subtracting the field equation for $\vec{c}_k .\vec{\varphi}_{\perp}$
from that for $\vec{c}_j .\vec{\varphi}_{\perp}$ then yields 
the equation of motion for $y_{jk}$: 
\begin{eqnarray}
\label{yequation}
\nabla^2 y_{jk} - \frac{M^2}{n} e^{-c \varphi}
 \left( \vec{c}_j .\vec{c}_j -\vec{c}_j .\vec{c}_k
\right) e^{-(\vec{c}_k.\vec{\varphi}_{\perp} +
\vec{c}_j .\vec{\varphi}_{\perp})/2}  \nonumber \\
\times
 \left[ \sinh (y_{jk}/2) \right] =0 .
\end{eqnarray}
The second term in Eq.~(\ref{yequation}) may be interpreted as the
derivative of an effective potential for $y_{jk}$.  The only critical
point in this potential is the global minimum that exists at
$y_{jk}=0$.  This implies that $y_{jk} \rightarrow 0$ at late times
and, consequently, $\vec{c}_j .\vec{\varphi}_{\perp} \rightarrow
\vec{c}_k . \vec{\varphi}_{\perp}$ {\em for all j and k}. 
However, taking the dot product of Eq.~(\ref{definen}) with
$\vec{\varphi}_{\perp}$ implies that $\sum_{i=1}^n \vec{c}_i
.\vec{\varphi}_{\perp} =0$ and consequently the $\vec{c}_i
.\vec{\varphi}_{\perp}$ can only be equal if they simultaneously
vanish: $\vec{c}_i . \vec{\varphi}_{\perp} =0$.  This condition is
identical (c.f. Eq.~(\ref{atcond})) to that placed on the solution
found above. We may conclude, therefore, 
that the solution given by
Eqs.~(\ref{fieldvarphi})--(\ref{fieldvec}) is the {\em unique}
late--time attractor for the cosmologies derived from Eq.~(\ref{4}).

The analytical form of the late--time attractor for the 
spatially flat, FRW cosmology is known 
when $c^2 \le 3$ and is given by the power law 
$a \propto t^{1/c^2}$ \cite{steep,liddle,exppot}. 
The self--interacting scalar field, $\varphi$, dominates
the massless fields, $\vec{\varphi}_{\perp}$, since 
the latter behave collectively as a stiff perfect fluid. 
Eq.~(\ref{cM}) implies that $c^2 \le  
3$ for an arbitrary dimensionality and the unique late--time 
attractor for the dimensionally reduced, spatially flat FRW
cosmology is therefore given by $a \propto t^r$,
where
\begin{equation}
\label{powerD}
r=1-\frac{2}{D-2} \,.
\end{equation}
In the five--dimensional model, where only one dimension is
compactified, the solution corresponds to that of a massless scalar
field, $r=1/3$. As more dimensions are compactified, the assisted
dynamics between the scalar fields becomes apparent and the power of
the cosmological expansion increases monotonically with the
dimensionality of spacetime. We can conclude from Eq.~(\ref{powerD}),
however, that assisted inflation does {\em not} arise in this
compactification scheme regardless of the number of compactified
dimensions, since as $D \rightarrow \infty$, $r \rightarrow 1$. It is
interesting that this upper limit corresponds precisely to the coasting
solution, where the scale factor is neither accelerating nor
decelerating. In summary, each compactification 
of an axion field effectively introduces
a new mass parameter, but the corresponding increase in the relative
expansion rate of the universe 
is counter--balanced by the new interactions that also
arise between the scalar dilaton fields \cite{cmn,ko}.

On the other hand, this model highlights a potential `moduli' problem
for early universe cosmology resulting from assisted scalar field
dynamics. In spatially flat FRW cosmologies containing a single scalar
field and a barotropic fluid with an equation of state $p=(\gamma -1
)\rho$, the late--time attractor is given by $a \propto t^{(1/c^2)}$
when $c^2 < 3\gamma /2$ \cite{coplidwan}.  When $c^2 > 3\gamma /2$,
there exists a scaling solution, where the energy density of the fluid
and scalar field redshift at the same rate
\cite{exppot,coplidwan,exppot1}.  An interesting case of
Eq.~(\ref{powerD}) corresponds to the six--dimensional model, where
the assisted dynamics between the scalar fields mimics the behaviour
of a relativistic fluid $(r=1/2, \gamma =4/3$). For $D\ge 7$, however,
$c^2 < 2$ and the scalar fields {\em dominate} the radiation
component.  This includes the compactified model derived from the
vacuum limit of M--theory.  In particular, when $D=8$ the expansion
rate is equivalent to that of a universe dominated by pressureless
dust. In this region of parameter space, therefore, primordial
nucleosynthesis may be significantly affected or may not proceed at
all unless the scalar field domination is reversed at a sufficiently
early epoch.  Consequently, this leads to limits on the duration of
scalar field domination in such models.

We may compare our results to those of Copeland, Mazumdar and Nunes
\cite{cmn}, who have studied the action (\ref{4}) when restricted to
the spatially flat FRW cosmology.  These authors found a power law
solution, $a \propto t^r$ $(r > 1/3)$, where the exponent is given by
\begin{equation}
\label{copelandpower}
r \equiv \sum_{i,j=1}^{n} \left( M^{-1} \right)_{ij} .
\end{equation}
Since $M_{ij}$ is non--singular, Eq.~(\ref{cM})
implies that $\sum_{i=1}^n (M^{-1})_{ij} =1/(c^2n)$. 
Summing over $j$ then implies that Eqs.~(\ref{cM}) and 
(\ref{copelandpower}) are equivalent. 

The above analysis may be readily extended to include the more 
general class of models with $n$ exponential 
potentials satisfying the simultaneous
constraints $\vec{c}_i.\vec{c}_i =\alpha + \beta$ and
$\vec{c}_i.\vec{c}_j =\beta$ $(i \ne j)$ for arbitrary constants
$\alpha$ and $\beta$.  Eq.~(\ref{defineM}) then implies that
$M_{ij} = \alpha \delta_{ij} + \beta$ and  
from Eq.~(\ref{cM}), this model 
admits a power law solution, where the exponent is given by 
$r = n/( \alpha + \beta n)$. 
The condition for inflation may therefore be deduced directly 
from the form of the coupling parameters. 

In conclusion, we have found a wide class of cosmological solutions
arising from a generalized Scherk--Schwarz compactification of vacuum,
Einstein gravity.  If only one potential was present, each would be
sufficiently steep for the moduli fields to track a radiative
fluid. However, the assisted dynamics between the fields implies that
they dominate the radiation.  Moreover, since the assisted dynamics is
insufficient to result in an inflationary expansion, regardless of the
dimensionality of the higher--dimensional spacetime, this leads to a
potential moduli problem for the early universe.

The power law solution of Ref. \cite{cmn} arises in the special case
of the spatially flat FRW cosmology. We have shown that this solution
is the unique late--time attractor. Since our analysis is independent
of the spacetime geometry, it can be employed to study multi--field
cosmology in the more general setting of spatially anisotropic and
inhomogeneous models. For example, particular solutions for a single
scalar field in a class of inhomogeneous Einstein--Rosen cosmologies
have been found previously \cite{G2}.  It would be interesting to make
a detailed study of the nature of the late--time attractors in these
models.

\acknowledgments

AMG is supported by the Particle Physics and Astronomy Research Council
(PPARC) UK. JEL is supported by the Royal Society. We thank 
E. J. Copeland, N. Kaloper, N. J. Nunes and D. Wands for helpful discussions.

\end{document}